\documentclass{article}

\usepackage{arxiv}

\usepackage[utf8]{inputenc} % allow utf-8 input
\usepackage[T1]{fontenc}    % use 8-bit T1 fonts
\usepackage{hyperref}       % hyperlinks
\usepackage{url}            % simple URL typesetting
\usepackage{booktabs}       % professional-quality tables
\usepackage{amsfonts}       % blackboard math symbols
\usepackage{nicefrac}       % compact symbols for 1/2, etc.
\usepackage{microtype}      % microtypography
\usepackage{lipsum}		% Can be removed after putting your text content
\usepackage{graphicx}
\usepackage{natbib}
\usepackage{doi}
\usepackage{amsmath}
\usepackage{bbm}

\title{Tail-Greedy Unbalanced Haar Wavelet Segmentation for Copy Number Alteration Data}

\author{
\includegraphics[scale=0.06]{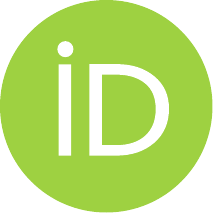}\hspace{1mm}Maharani Ahsani Ummi \\
Statistics Research Division\\
Bandung Institute of Technology\\
Jl. Ganesa No.10, Bandung, West Java 40132, Indonesia \\
\texttt{maharaniahsani@itb.ac.id} \\
\And
\includegraphics[scale=0.06]{orcid.pdf}\hspace{1mm}Stuart Barber \\
Department of Statistics\\
University of Leeds\\
Woodhouse Lane, Leeds, West Yorkshire LS2 9JT, United Kingdom \\
\texttt{} \\
\And
\includegraphics[scale=0.06]{orcid.pdf}\hspace{1mm}Henry M. Wood \\
Leeds Institute of Medical Research at St. James's\\
University of Leeds\\
Leeds, West Yorkshire LS2 9JT, United Kingdom \\
\texttt{} \\
\And
\includegraphics[scale=0.06]{orcid.pdf}\hspace{1mm}Arief Gusnanto \\
Department of Statistics\\
University of Leeds\\
Woodhouse Lane, Leeds, West Yorkshire LS2 9JT, United Kingdom \\
\texttt{} \\
}

\renewcommand{\shorttitle}{\textit{arXiv} Template}

\hypersetup{
pdftitle={A template for the arxiv style},
pdfsubject={q-bio.NC, q-bio.QM},
pdfauthor={David S.~Hippocampus, Elias D.~Striatum},
pdfkeywords={First keyword, Second keyword, More},
}

\begin{document}
\maketitle

\begin{abstract}
Detecting copy number alterations (CNAs) from next-generation sequencing data remains challenging, particularly for short segments under noisy conditions. Existing segmentation methods often suffer from high false positive rates or fail to reliably detect short aberrations, especially in low-coverage data. In this study, we propose a modified tail-greedy unbalanced Haar (TGUHm) method that introduces a dual-thresholding strategy to improve segmentation accuracy. The proposed approach effectively suppresses spurious spikes while preserving sensitivity to both short and long CNA segments. Extensive simulation studies under Gaussian and heavy-tailed noise demonstrate that TGUHm consistently achieves higher true positive rates and lower false positive rates compared to state-of-the-art methods, including CBS, HaarSeg, and FDRSeg. In particular, the proposed method improves detection accuracy for short segments while maintaining competitive overall performance. Application to real cancer genomic data further confirms the practical utility of the method, revealing biologically meaningful CNAs associated with known cancer-related genes. These results suggest that TGUHm provides a robust and effective framework for CNA detection in challenging sequencing settings.
\end{abstract}

% keywords can be removed

\section{Introduction}
\label{Introduction}
Recent technological development of next-generation sequencing (NGS) has transformed the way we investigate genomic structural variation such as copy number alteration (CNA). These alterations mean that some genomic regions exhibit more (gains) or less (losses) copy number than the normal two copies; they are identified as `jumps' and `drops', respectively, in the copy number ratio \citep{Gusnanto2012}.  The CNA of any genomic region can be estimated from the ratio of the number of short segments (`reads') aligned to the region between tumour and normal samples \citep{Gusnanto2012}. The ratios are calculated for each non-overlapping fixed-size `window' or `bin' along the genome. The ratios are inherently noisy due to random experimental error between the paired samples, and an important preprocessing step called segmentation is needed to deal with the noise in the analysis of CNA data. The goal of this segmentation step is to remove the random error and to recover the unknown function that represents the true copy number changes on the chromosome from the observed ratios, so that subsequent downstream analyses can be carried out based on these information. This underlying pattern can be considered to be a piecewise constant function and, therefore, we can say that the identification of CNA pattern is, in principle, a change-point detection and estimation problem.

To deal with this problem, several authors have proposed change-point detection methods for segmentation. For example, \cite{Olshen2004} proposed a method called circular binary segmentation (CBS) that identifies the change-points by comparing a maximal t-statistic to a permutation reference distribution to obtain the corresponding $p$-value. This is followed by a pruning algorithm to control the number of change-points. Another approach was proposed by \cite{Muggeo2010} that utilized a simple transformation based on the cumulative sum of the data. \cite{BenYaacov2008} proposed a wavelet-based approach to segmentation, where a traditional non-decimated Haar wavelet transform was investigated to estimate CNA. In a more recent approach, a multiscale segmentation method controlling the false discovery rate (FDR) was proposed by  \cite{Housen2016} and a least-squared based segmentation algorithm combining least-squares principles and a suitable penalization scheme to perform copy number segmentation was introduced by \cite{Nilsen2012}. While other segmentation methods exist, some comparison studies, such as in \cite{Lai2005} and \cite{Willenbrock2005} which tested several segmentation methods, concluded that CBS tend to have the best performance under various conditions. 

In the context of low-coverage \cite{Wood2010} ($<$ 0.1X), many of these methods are well known to work well in detecting long segments (say, $>10$ Mbp) but are not sensitive to short segments (approximately 1 Mbp), and often unable to detect them. However, detection of short segments remains a major challenge since such segments are represented by only a few `windows' or points in the low-coverage setting \citep{Gusnanto2014}.  In cancer research, short segments potentially contain key oncogenes or tumour-suppressor genes of interest, so a method that performs well in estimating both long and short segments is generally needed.  In short segments, the underlying true change is masked by noise
% error variability with complex distribution
and it is very difficult to distinguish between the two. To address this challenge, we consider a wavelet segmentation approach. The main reason behind this choice of approach is that it enables careful investigation of variability at different scales. The latter is important in our context because we wish to detect real short-term changes that are close in magnitude to the error variability (in addition to longer segments). Therefore, this approach gives us a major tool for identifying (and dealing with) segments on a range of scales.

In this study, we present an approach called TGUHm method which applies the tail-greedy unbalanced Haar (TGUH) wavelet transform to perform segmentation of CNA data. The TGUH was originally proposed by \cite{Fryzlewicz2018} for a one-dimensional signal plus noise model.  While a traditional `balanced' Haar wavelet is a step function that changes value at the mid-point of its support, the `unbalanced' aspect of the TGUH approach allows the change-point in the basis functions to occur anywhere in the interior of each wavelet's support, and indeed to be in different locations on different basis functions. This characteristic gives the benefit that the data length does not have to be a power of two as in the traditional discrete Haar wavelet method. An additional benefit is it can take in to avoid the problem that commonly arises in Haar wavelet estimation where the estimator is more likely to detect jumps at dyadic locations, i.e, $1/2, 1/4, 3/4, \ldots,$ which might not be the actual locations of the jumps/drops in the true underlying CNA pattern. 

While providing an appealing ability in estimating correct change-point locations, an extreme observation commonly found in NGS data will tend to cause TGUH method to detect spikes which are likely to be estimated as spurious change-points as a result of the `bottom-up' approach used in the TGUH transform. To address this problem, in our TGUHm method, we introduce a simple thresholding procedure to control the minimum altered segment size which is applied following the connected thresholding used in the original TGUH method. This modification improves the performance of the TGUH method. Our investigation in this study suggests that this approach gives good operating characteristics in detecting both short and longer segments. We also illustrate the application of this method on a real dataset from a lung cancer cohort where the CNA data are obtained from NGS with a low-coverage setting.

\begin{figure*}[h]
	\begin{center}
		\includegraphics[scale=0.42]{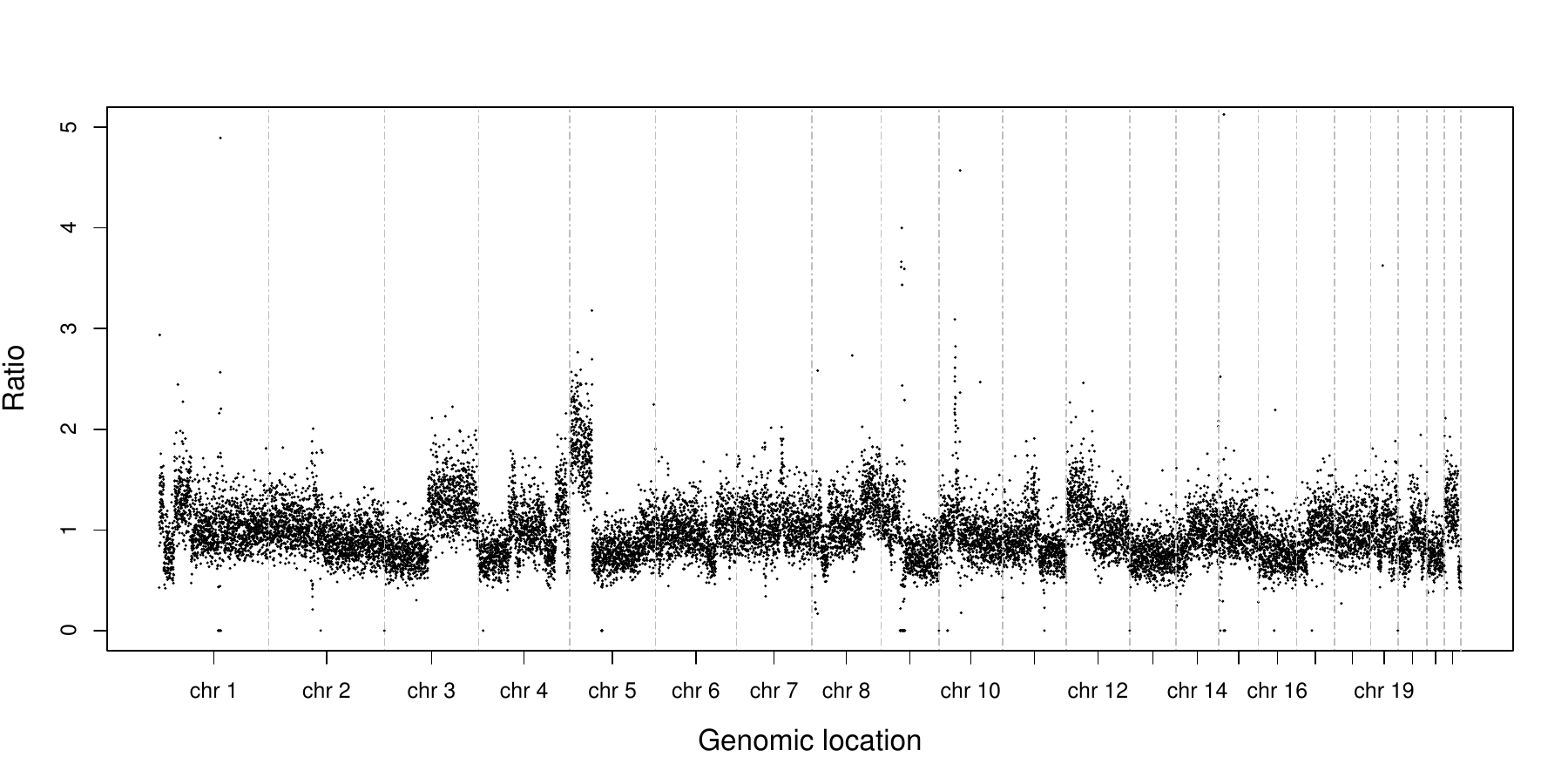}
	\end{center}
	\caption{Example of copy number ratio data from one patient, TMA-93. The data was normalised using CNAnorm \cite{Gusnanto2012} and regions with missing values, such as the centromeres, are removed. Each point in the figure denotes the copy number ratio of TMA-93 which corresponds to a specific genomic window (150 kb).}
	\label{fig:TMA-93}
\end{figure*}

\section{Method }\label{sec:method}

\subsection{Patients, sequence data and alignments}

In this study, we considered squamous cell carcinoma lung cancer samples from \cite{Belvedere2012}. Details on sample preparation, DNA extraction and library preparation are described by \cite{Wood2010}. Sequences were aligned using the bwa suite version 0.5.9-r16 \citep{Li2009} against assembly hg19 of the human genome. Only sequences that could be uniquely aligned and with mapping quality $\geq 37$ were used. Using `depth of coverage' we counted the short sequences (`reads') that were mapped to fixed non-overlapping genomic regions (`windows'), estimated to be 150 kbp \citep{Gusnanto2014}. Although the proposed segmentation method in this paper can be implemented to either non-normalised or normalised data, we performed a normalisation using the CNAnorm package \citep{Gusnanto2012} for easier comparison between CNA profiles. Regions with missing values, such as the centromeres, are removed. An example of CNA data from patient TMA-93 is shown in Figure~\ref{fig:TMA-93}. Each point in Figure~\ref{fig:TMA-93} denotes the copy number ratio of patient TMA-93 corresponding to a specific genomic window (with size 150 kb). To demonstrate the generalisability of the proposed method to data from different sequencing technology, we also consider the breast cancer dataset from \cite{Snijders2001}. 

\subsection{Tail-greedy unbalanced Haar wavelet segmentation for NGS data}

Before describing the tail-greedy unbalanced Haar wavelet method, we define some relevant notation and concepts. Let $n$ be the number of windows/regions; we segment each chromosome separately, so $n$ is the number of windows in a chromosome and we do not need an index to denote chromosome.  Alternatively, one can segment on the whole genome simultaneously, in which case $n$ would denote the number of windows in the entire genome. Let $x_i$ denote the  location of the $i$-th window in the chromosome/genome for $i=1,2, \dots ,n$, satisfying the condition $x_1 < x_2 < \dots < x_n$. Let $N$ be the number of change-points in the data, with $0 \leq N \ll n$, and if $N>0$, let $\eta_j$, $j=1, \dots,N$ be the locations of the change-points. For a sequence $\{y_i\}_{i=1,\ldots,n}$, we say a change-point is located at $\eta_j = x_k$ if $|y_{k+1}-y_k|>\theta$, where $0 < k \leq n$ and the height tolerance parameter $\theta$ is set to be equal to $0.1$ \citep{Mermel2011}. As an illustration, $\eta_2=x_{100}$ means that the second change-point in the data is located at $x_{100}$ and $|y_{101}-y_{100}|>\theta$. In a simulation study, $N$ and the $\eta_j$'s are known, but in practice for real data they are unknown. 

In the context of NGS, let $y_i$ denote the ratio between the number of reads in the tumour and normal sample in the $i$-th window corresponding to location $x_i$ \citep{Gusnanto2012}.  There is no requirement for the $y_i$ to be normalized. In the normalization of CNA data from clinical samples, segmentation may be involved at the start and the end of normalization \citep{Gusnanto2012}. To accommodate other normalization methods that users may use, we assume/consider the proposed segmentation method to estimate CNA from normalized ratios. %In the context aCGH data, we let $y_i$ denote the log ratio of intensities between the tumour and normal sample in the $i$-th genomic region at the corresponding location $x_i$.

We assume that the observed $y_i$ are given by a true (unknown) signal $f_i$ obscured by additive random error. This model can be expressed as
\begin{align}
	y_i = f_i + \epsilon_i, \label{eq:main_model}
\end{align}
where $\epsilon_i$ represents measurement noise and $f$ is a one-dimensional, piecewise-constant signal with change-points at unknown locations $\eta_1, \ldots , \eta_N$. 

From a statistical perspective, this problem can be viewed as the recovery of a piecewise-constant signal $f$ from the noisy data $y$, where reliable detection of change-points depends on distinguishing true structural changes from random fluctuations. In the context of wavelet-based methods, small-scale coefficients are typically associated with noise, while larger-scale coefficients capture meaningful signal variations. However, in the presence of extreme observations or heavy-tailed noise, small-scale coefficients may occasionally attain large values, leading to spurious detections.

The proposed modification addresses this issue by introducing an additional thresholding mechanism that explicitly constrains the minimum segment size. This can be interpreted as a form of regularization that penalizes overly localized changes, which are more likely to arise from noise rather than true signal. By combining the hierarchical structure of connected thresholding with this additional constraint, the method effectively balances sensitivity to genuine short segments and robustness against noise-induced artifacts. As a result, the proposed approach improves the stability of the segmentation while preserving its ability to detect meaningful changes across multiple scales.

To operationalize this idea, we build upon the standard Tail-Greedy Unbalanced Haar (TGUH) framework and incorporate the proposed modification within its thresholding procedure. The standard TGUH approach proposed by \cite{Fryzlewicz2018} consists of three main steps: (i) Forward TGUH transform, (ii) Thresholding and (iii) Inverse TGUH transform. 

The proposed TGUHm method follows the general structure of the TGUH algorithm, consisting of three main steps. However, key modifications are introduced in the thresholding and reconstruction stages to improve robustness against noise and reduce spurious detections. The overall procedure is outlined below.
% In this study, we propose a method called TGUHm that mainly follows these steps but with some modifications in steps (ii) and (iii) to adapt to the characteristics of CNA data from NGS. We now briefly outline these steps for the proposed TGUHm method, before subsequently explaining them in detail.
\begin{enumerate}
	\item Apply the (standard) TGUH transformation to the data to get TGUH detail coefficients, which give a sparse representation of the data in terms of a set of piecewise-constant basis functions. The coefficients are assigned into a unary-binary tree (i.e., one in which each `parent' coefficient has one or two	`child' coefficients). 
	\item Threshold or delete those detail coefficients whose values are less than a specified threshold. Two-stage thresholding is performed here to firstly remove smaller coefficients that are believed to represent noise $\epsilon$ rather than the true signal $f$. This is part of the standard TGUH method. Additionally, in the second stage, we delete some coefficients that correspond to `spikes' that may occur due to extreme outliers in copy number ratios. 
	\item Reconstruct the segmentation result by returning the sample mean of the observed data within each segment between consecutive estimated change-points. This step is different to that in the standard TGUH method.
\end{enumerate}

\subsubsection{Step 1: TGUH transformation \label{sec:transform}} 
The TGUH transformation is a bottom-up method which, in each iteration, selects adjacent pairs of the data which are thought to have the smallest variability in an attempt to concentrate as little as possible of the variability or `power' in the data at the `finer' or lower levels of resolution \citep{Fryzlewicz2018}.  Let $c_{s,e}$ be the local rescaled average of copy number ratio data $y_i$ in the region $i=s, \ldots, e$, given by
\begin{equation}
	c_{s,e} = \frac{1}{\sqrt{e-s+1}} \sum_{i=s}^{e} y_i.
	\label{eq:cse}
\end{equation}
The two subscripts in $c_{s,e}$ denote the start ($s$) anda end ($e$) index of the subset of the data which corresponds to $c_{s,e}$. For example, $c_{1,1}$ corresponds to $y_1$, $c_{2,3}$ corresponds to $y_2$ and $y_3$. At each iteration, we select pairs of consecutive regions and merge them by applying an orthonormal transformation. Define $C^j=\{c_{s,e}\}$ to be the set of local rescaled average coefficients of the data $y_i$, and $\alpha_{j}$ to be the regions remaining, after the $j$-th iteration. For $j=1$, we assign $C^1 = \{c_{1,1},c_{2,2},\ldots,c_{i,i}\} =\{y_1,y_2,\ldots,y_i\}$. Let $\rho$ be a parameter describing the proportion of pairs to merge in each iteration; we merge $\lceil \rho \alpha_{j} \rceil$ pairs at the $j$-th iteration. In our applications, we use $\rho = 0.01$.

To be more precise, the algorithm proceeds as follows:
\begin{enumerate}
	\item At the $j$-th iteration, for each adjacent pair of local rescaled average coefficients, construct a `detail' filter $(l_{s,b},-r_{b+1,e})$ with $l_{s,b}^2+r_{b+1,e}^2=1$ and $d_{s,b,e}= l_{s,b}c_{s,b} - r_{b+1,e}c_{b+1,e}$ should be zero if $(y_s,\ldots,y_e)$ is a constant vector.  
	\item Compute the detail coefficient $d_{s,b,e}= l_{s,b}c_{s,b} - r_{b+1,e}c_{b+1,e}$ for each adjacent pair of coefficients in $C^j$ and sort the sequence $| d_{s,b,e} |$ in ascending order. Then search for the $\lceil \rho\alpha_{k} \rceil $ pairs of local rescaled average coefficients that have the smallest absolute value of the detail coefficient vector and save them as detail coefficients of scale $j$, $d^{j,k}_{s,b,e}$, where $k=1, \ldots, K(j)$ denote the index of them according to tincreasing $s$.
	\item Merge the local rescaled average coefficients which correspond to the selected detail coefficients. Then, produce new local rescaled average coefficients which define the scaled average of those merged regions. Specifically, for a selected detail coefficient $d^{j,k}_{s,b,e}$  at iteration $j$, we merge the regions $\{s, \ldots, b\}$ and $ \{b+1, \ldots, e\}$ into single region $\{s, \ldots , e\}$ and produce new local rescaled average coefficients $c_{s,e}$. After this step, we will have a new set of local rescale average coefficients $C^{j+1}$.
	\item Set $j = j+1$ and go back to step (1), unless only one detail coefficient was extracted in step 2, in which case the algorithm terminates.
\end{enumerate}
This step provides a multiscale representation of the data, enabling the identification of changes at different resolutions.

\subsubsection{Step 2: Thresholding \label{sec:thresholding}}
This step is the key component of the proposed method, where modifications are introduced to control noise-induced artifacts. In a wavelet context, thresholding is commonly used to remove noise from data by shrinking/deleting some wavelet coefficients that fall below a specified threshold. In the TGUH decomposition, by construction, the bulk of the activity of the data will be concentrated in coarse-scale (large $j$) detail coefficients and fine-scale (small $j$) coefficients will be small and contain mostly noise. Therefore, by removing those coefficients which are smaller than some threshold, we can remove much of the noise. But in some cases where there is a frequent occurrence of outliers, as is often found in NGS data, basic wavelet thresholding tends to overestimate the number of change-points, detecting spurious change-points as spikes (very short altered segments of only one data points). In our TGUHm method, we therefore, add an additional procedure to the connected thresholding used in TGUH \citep{Fryzlewicz2018} for pruning these spikes. Hence, the thresholding procedure in our proposed TGUHm method proceeds in the following two stages.
\begin{enumerate}
	\item \emph{Connected thresholding}. Perform connected thresholding to detail coefficients $d^{j,k}_{s,b,e}$. This thresholding is used by \cite{Fryzlewicz2018} which preserves the `unary-binary' structure of the detail coefficients and produces an estimate where the number of change-points is equal to the number of detail coefficients.
	
	Let the children coefficients of detail coefficient $d^{j,k}_{s,b,e}$ be the set of finer-scale coefficients whose support is entirely inside $[s,e]$:
	\begin{align}
		\mathcal{C}^j_{s,b,e} 
		&= \{d^{j',k'}_{s',b',e'}: [s',e'] \subseteq [s,e] \notag \\
		&\quad \text{for all } j' = 1, \ldots, j, \text{and } k'=1, \ldots, K(j') \}.
	\end{align}
	Connected thresholding, with threshold $\lambda>0$, sets to zero all detail coefficients $d^{j,k}_{s,b,e}$ for which  $|d^{j,k}_{s,b,e}|<\lambda$ and each of its children coefficients are also smaller in magnitude than $\lambda$. More formally, if $g^{j,k}_{s,b,e}$ and $d^{j,k}_{s,b,e}$ are the detail coefficients respectively of the true unknown signal $f$ and the observed data $y$ in Equation~(\ref{eq:main_model}), the connected thresholding estimate of $g^{j,k}_{s,b,e}$ is given by
	\begin{equation}
		\hat{g}^{j,k}_{s,b,e} = d^{j,k}_{s,b,e} \mathbbm{1}\{ \exists d^{j',k'}_{s',b',e'} \in \mathcal{C}^j_{s,b,e} > \lambda\},
	\end{equation}
	where $\mathbbm{1}\{\cdot\}$ is the indicator function.
	\item \emph{Unconnected thresholding}. We propose an additional `unconnected' form of thresholding after the above step, where we do not preserve the `unary-binary tree' structure of detail coefficients. This reduces the tendency of connected thresholding to leave `spikes' in the estimated segmentation. `Spikes' are likely to occur when the detail coefficients $d^k_{s,b,e}$, with either $e-b$ or $b-s+1$ equals to one, survive the thresholding. To control the occurrence of `spikes', we set to zero the detail coefficients $d^k_{s,b,e}$ with either $e-b$ or $b-s+1$ less than a constant $c^*$, say. By setting $c^*=2$, we could reduce spikes and have a direct control on the minimum length of segments. Our final estimator $\tilde{g}^k_{s,b,e}$ of $g^k_{s,b,e}$ is given by
	\begin{equation}
		\tilde{g}^{j,k}_{s,b,e} = \hat{g}^{j,k}_{s,b,e} \mathbbm{1}\{ (b-s+1)>c^*\} \mathbbm{1}\{(e-b)>c^* \}.
	\end{equation}
\end{enumerate}

The proposed two-stage thresholding procedure is designed to enhance both the sensitivity and specificity of feature detection. In the first stage, connected thresholding ensures that important features are not prematurely removed, particularly those that are only prominent at finer scales. This is achieved by preserving the hierarchical (unary-binary tree) structure of the detail coefficients, where a coefficient at a coarser scale is retained if any of its child coefficients at finer scales exceed the threshold. This hierarchical dependency prevents the loss of subtle but meaningful variations that might otherwise be discarded if evaluated in isolation. In the second stage, unconnected thresholding is applied to further refine the results by eliminating segments that are too small, thus reducing over-segmentation caused by noise. This is particularly important in genomic data analysis, where it is undesirable to flag regions with only one or two data points as significant alterations, as these are often artifacts rather than true biological signals. Together, these stages create a robust framework that balances the need to detect subtle features while minimizing false positives.

In our application, we define the threshold parameter $\lambda$ using $\lambda = \sigma(2(1+0.01) \log n)^{1/2}$ as in \citep{Fryzlewicz2018}. In practice the noise level parameter $\sigma$ is unknown, so we estimate it by computing the median absolute deviation (MAD) of the sequence $\{|y_{i+1}-y_{i}|/\sqrt{2}; i = 1, \ldots, n-1\}$; these values are the finest-scale balanced Haar wavelet coefficients of the vector $y$.

\subsubsection{Step 3: Signal reconstruction}

Unlike the original TGUH method of \cite{Fryzlewicz2018}, we do not reconstruct the segmentation result by performing the inverse TGUH transform. This is due to the additional unconnected thresholding used in the previous step. If we apply the inverse TGUH transform directly to the unconnected thresholding results, there may occur a situation where the estimated signal for a segment is not equal to the sample mean of the data in that segment. 

Since each breakpoint $b$ in the middle of surviving wavelet/detail coefficients $\tilde{g}^{k}_{s,b,e}$ denote the locations of change-points, we therefore, estimate the piecewise constant signal $f_i$ between two consecutive change-points by the sample mean of all copy number ratio data $y_i$ in that interval. 
More formally, let $\mathbf{b} = \{b_l\}$ denote the collection of $b \in \tilde{g}^{k}_{s,b,e}$ in ascending order where $l =1,..,N$ and we can say that $N$ is the number of estimated change-points. Define $\eta_j = \{\eta_1,\eta_2, \ldots, \eta_{N+2}\} =  \{0, b_1, b_2, \ldots, b_N, n\}$; $n$ is the length of the copy number ratio data $y$. So that, the final estimator $\hat{f}$ of the true function $f$ in Equation~(\ref{eq:main_model}) is defined by
\begin{equation}
	\hat{f}_j = \frac{1}{\eta_{j+1} - \eta_j} \sum_{k=\eta_j}^{\eta_{j+1}} y_k.
\end{equation}
This reconstruction ensures that the estimated signal reflects stable segment-level behavior rather than local fluctuations.

\subsection{Simulation study \label{sec:simulation}}
\begin{figure*}[h]
	\centerline{\includegraphics[scale=0.4]{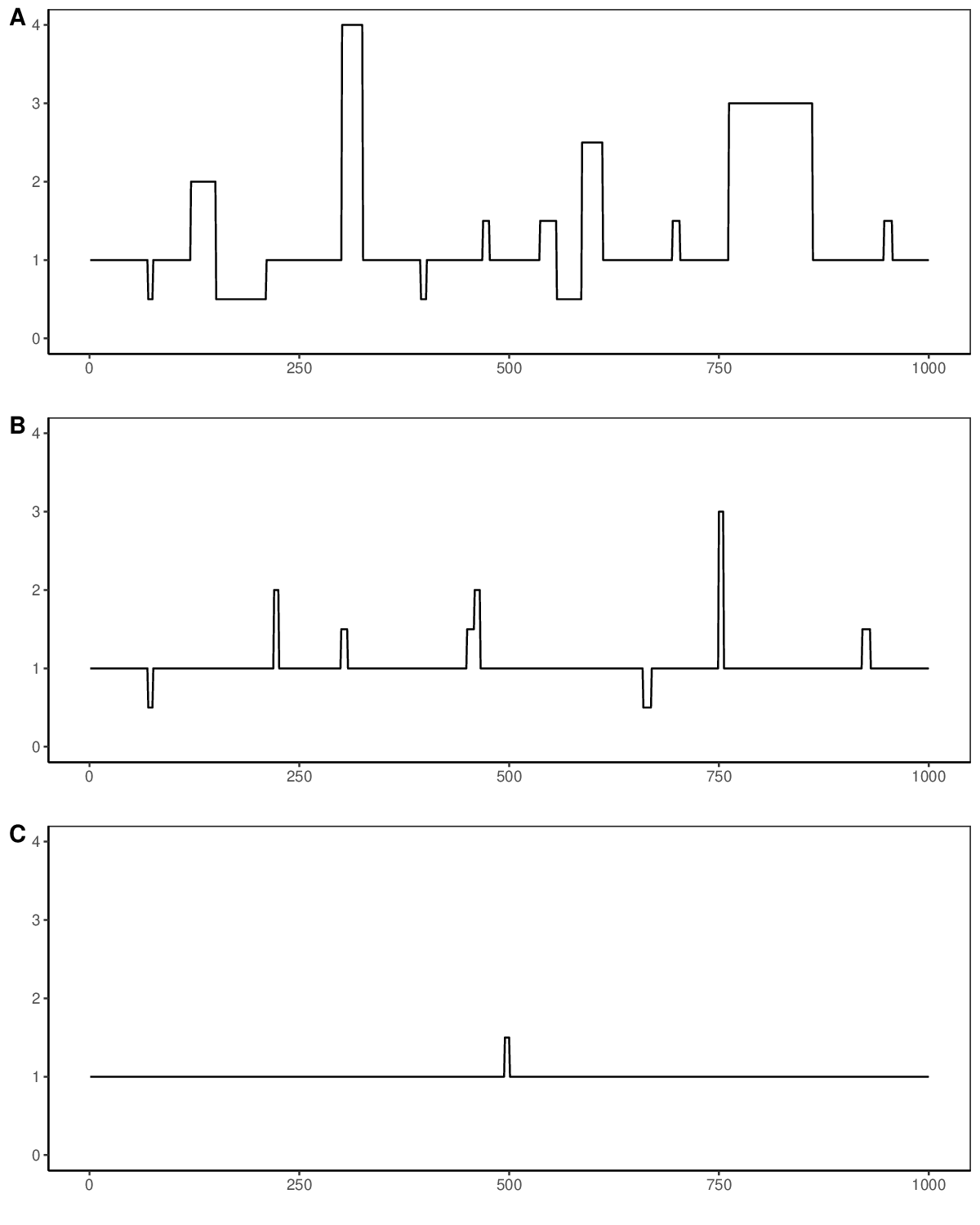}}		
	\caption{The true patterns of copy number alterations, denoted $f$, in simulated examples. \textbf{(A)} First true function. The irregular pattern of segment length is based on common patterns observed in real data. \textbf{(B)} Second true function, which aims to characterise the proposed method's performance in a case where the underlying true pattern only contains short altered segments. \textbf{(C)} Third true function. An extreme case where there is only a short altered segment in the middle of long segment.}
	\label{fig:test_obj}
\end{figure*}

To evaluate the performance of the proposed method, we conducted a comparative simulation study by considering three test functions shown in Figure~\ref{fig:test_obj}. The first true function is based on some patterns of different segment lengths including both long segments and short segments commonly observed in real data and the aberrations in height/depth varies between $0$--$4$. In this simulation study, we refer to aberrations with length between $6$--$10$ data points as short segments while long segments comprise more than $10$ data points. This is based on the window size used in our data (150kb), in which a 1 Mb segment is represented by only 6-7 windows or data points. The height of the short segments is set to $0.5$ to represent a typical smallest change that we might expect in real data. The second function only includes short segments with various heights to assess the ability of the method in estimating short segments. Lastly, for the third simulation, we consider an extreme case where there is only a single altered segment with a very short (6 point) width.

We generate 1000 replicates for each of the true functions and use two kinds of noise model to contaminate those data. The first noise model is i.i.d.\ Gaussian noise $N(0,\sigma^2)$ and the second is a heavier-tailed noise model that reflects extreme observations that often occur in NGS copy number ratio data. A classical way to generate the second noise model is to use ``contaminated normals'', where the error distribution is a mixture of two normal distributions \citep{Turkey1960}. With probability $1 - \alpha$ the error is drawn from a distribution $N(0, \sigma^2)$, and with probability $\alpha$ from $N(0, d^2\sigma^2)$, with $d = 3$ and $\alpha = 0.05$ \citep{Nilsen2012}. We repeat our simulations for $\sigma = 0.1, \ldots, 0.5$ for both noise to obtain a controlled comparison of different levels of noise variance relative to the changes that we wish to detect in CNA data, which are generally of magnitude $0.5$ or $1$ \citep{Gusnanto2012}. 

Since we know where the true change points are located in simulated data, we can consider the problem of change points detection as a binary classification problem. Correctly identified change-points (true positive, TP) are those whose locations are found within two windows and closest to the true change point. If there are two closest change-points detected, we assign one as TP and the other one as false positive (FP). The remaining change points detected, $\text{FP}=\text{P}-\text{TP}$, where P denotes positives or the total number of estimated change points $N$, are considered spurious estimates (FP) \citep{Pierre2015}. The illustration of these definitions is presented in Figure~S1 of the Supplementary Material. Based on this definition we compute the average of true positive rate (aTPR) and the average false positive rate (aFPR) over 1000 replicates. To assess the ability of the method in estimating short segments, we also calculate the average true positive rate in estimating short segments (aTPRsh).

To further evaluate the operating characteristics of each method, we also calculate the Receiver Operating Characteristic (ROC) curve for each method across different values of $\sigma^2$. The ROC curves are plotted based on the mean TPR and FPR across replicate data sets for each segmentation method.

For all the simulations above, we performed TGUHm and compare its performance to the original TGUH method \citep{Fryzlewicz2018}. The R code of TGUHm method is available at {\sf https://github.com/maharaniau/TGUHm}. To evaluate the practical impact of the constant $c^*$, we compare $c^*=1$ and $c^*=2$. For $c^*=1$, both TGUHm and TGUH will produce exactly the same results, which we denote by TGUH1. Besides the original TGUH method, we also performed TGUH combined with a localised pruning method using the {\sf R} package {\sf breakfast ver 2.2} which we denote by TGUHb.  We also consider several recently published methods as competitors: Circular Binary Segmentation (CBS) \citep{Olshen2004}, HaarSeg \citep{BenYaacov2008} which is based on the balanced Haar wavelet transform, CopyNumber  \citep{Nilsen2012}, CumSeg \citep{Muggeo2010}, and FDRseg \citep{Housen2016}. We apply the CopyNumber method twice, with its main parameter $\gamma$ set to be $12$ and $40$ as suggested by \cite{Nilsen2012} to give different balances between sensitivity and specificity. The results for these two separate analyses are denoted as Copy12 and Copy40, respectively.

\section{Results}

\subsection{Simulation result}
\begin{figure*}[p]
	\begin{center}
		\includegraphics[scale=0.35]{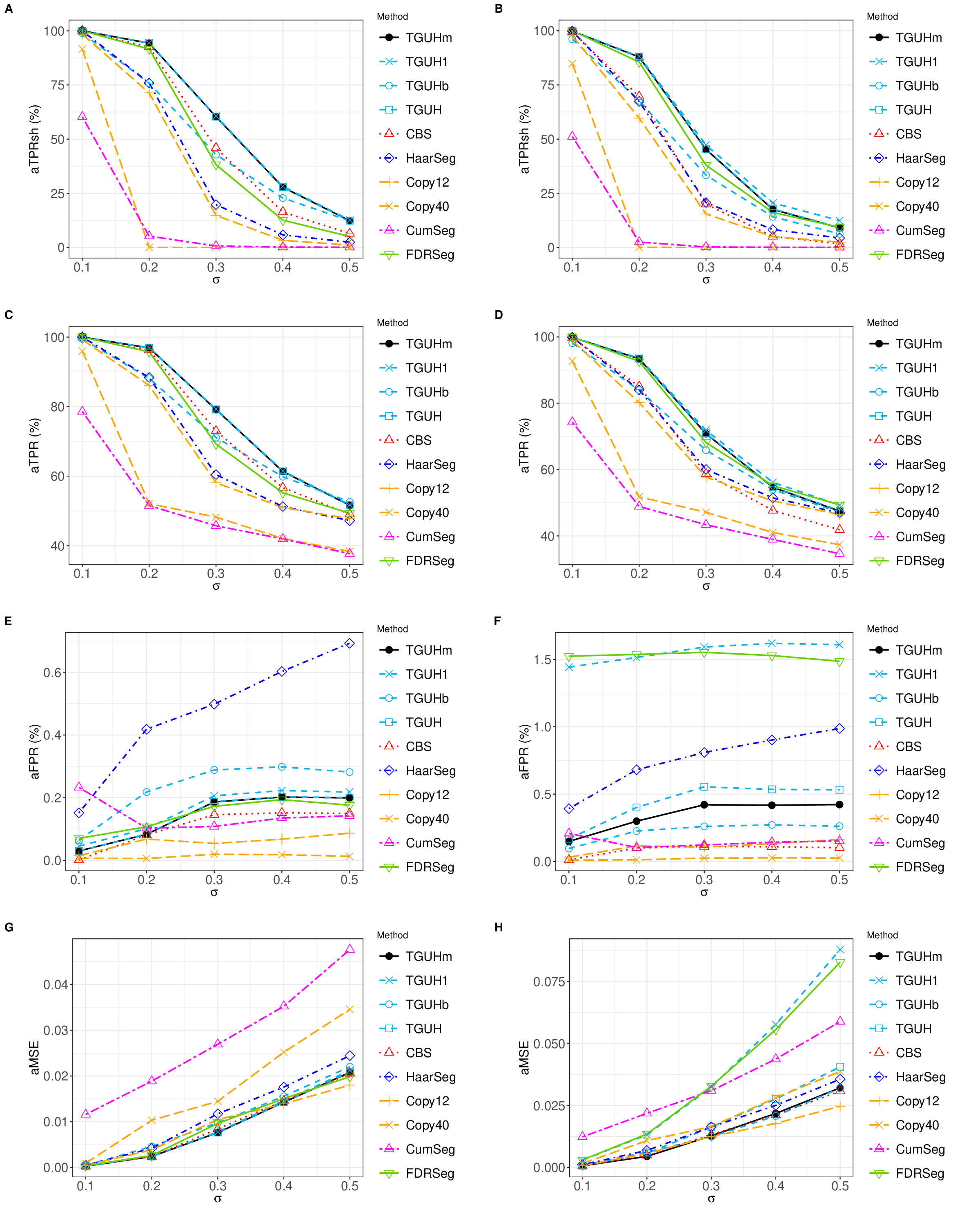}
	\end{center}
	\caption{Performance metrics for 1000 replicates of the first test function (see panel \textbf{A} of Figure~\ref{fig:test_obj}).  \textbf{(A) (B)} Average of true positive rate in estimating change-points that corresponds to short segments (aTPRsh). \textbf{(C) (D)} Average true positive rate (aTPR). \textbf{(E) (F)} Average of false positive rate (aFPR). \textbf{(G) (H)} Average of mean-square error (aMSE) of the estimated piecewise constant signal to the true function.  The left column (panels \textbf{A},\textbf{C}, \textbf{E}, and \textbf{F}) show results for  i.i.d.\ Gaussian noise $N(0,\sigma^2)$, while the right column (panels \textbf{B},\textbf{D}, \textbf{F}, and \textbf{H}) show results for noise from a mixture of two Gaussian distributions $0.95\times N(0,\sigma^2)+0.05 \times N(0,3\sigma^2)$). For a quick reminder, TGUH1 denotes both TGUH and TGUHm method with $c^*=1$ while TGUHm and TGUH denote TGUHm and TGUH method with $c^*=2$, respectively. TGUHb denotes TGUH method with a localised pruning algorithm. Copy12 and Copy40 denote CopyNumber method with $\gamma$ parameter equal to 12 and 40, respectively.
	}
	\label{fig:type_8}
\end{figure*}

\begin{figure*}[p]
	\begin{center}
		\includegraphics[scale=0.35]{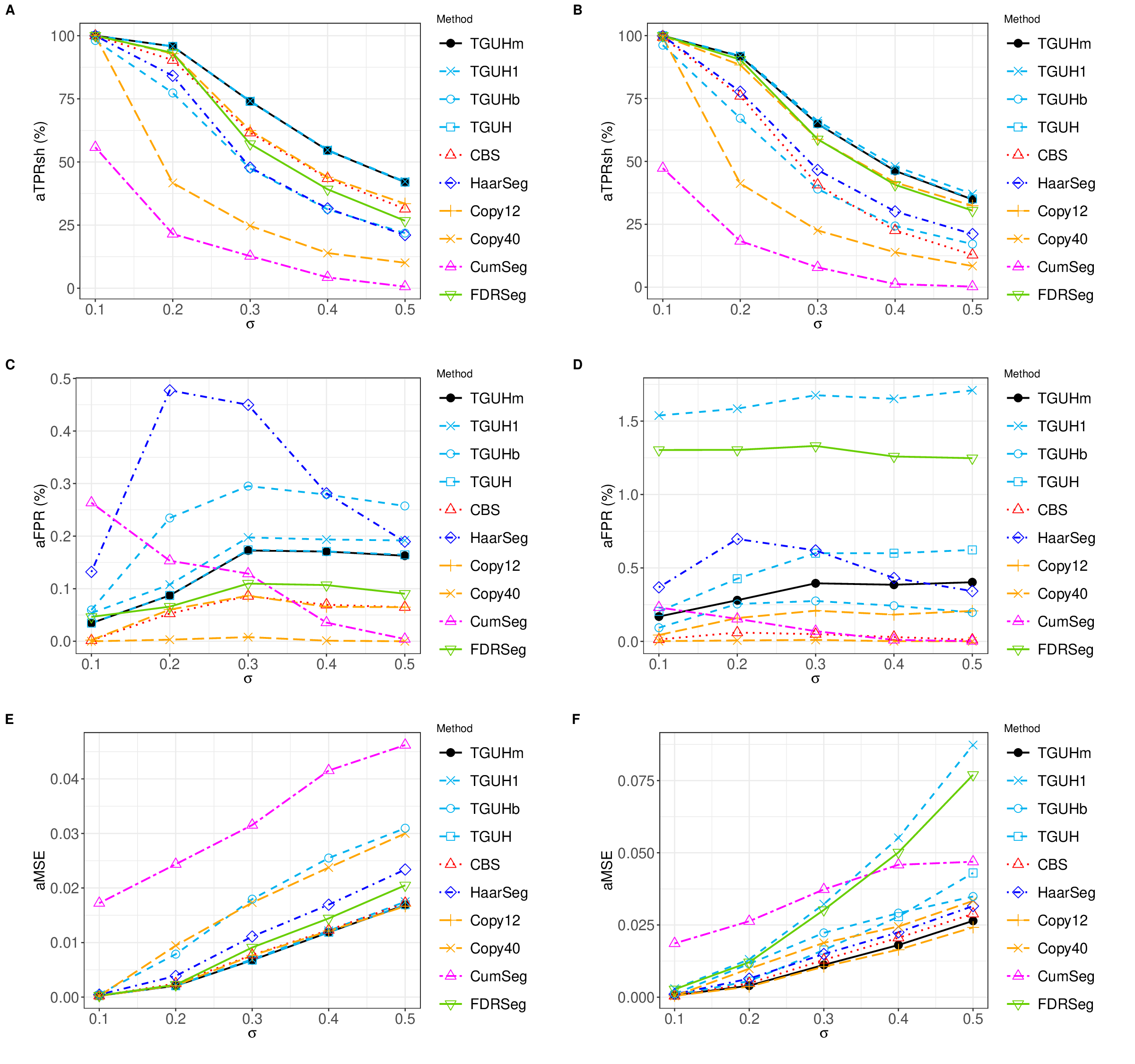}
	\end{center}
	\caption{Performance metrics for 1000 replicates of the second test function (see panel \textbf{B} of Figure~\ref{fig:test_obj}). \textbf{(A) (B)} Average of true positive rate in estimating change-points that corresponds to short segments (aTPRsh). \textbf{(C) (D)} Average of false positive rate (aFPR). \textbf{(E) (F)} Average of mean-square error (aMSE) of the estimated piecewise constant signal to the true function. The left column (panels \textbf{A},\textbf{C}, and \textbf{E}) show results for  i.i.d.\ Gaussian noise $N(0,\sigma^2)$, while the right column (panels \textbf{B},\textbf{D}, and \textbf{F}) show results for noise from a mixture of two Gaussian distributions $0.95\times N(0,\sigma^2)+0.05 \times N(0,3\sigma^2)$). The aTPR results are omitted as the simulated data only contains short segments. For a quick reminder, TGUH1 denotes both TGUH and TGUHm method with $c^*=1$ while TGUHm and TGUH denote TGUHm and TGUH method with $c^*=2$, respectively. TGUHb denotes TGUH method with a localised pruning algorithm. Copy12 and Copy40 denote CopyNumber method with $\gamma$ parameter equal to 12 and 40, respectively.
	}
	\label{fig:type_9}
\end{figure*}

\begin{figure*}[p]
	\begin{center}
		\includegraphics[scale=0.35]{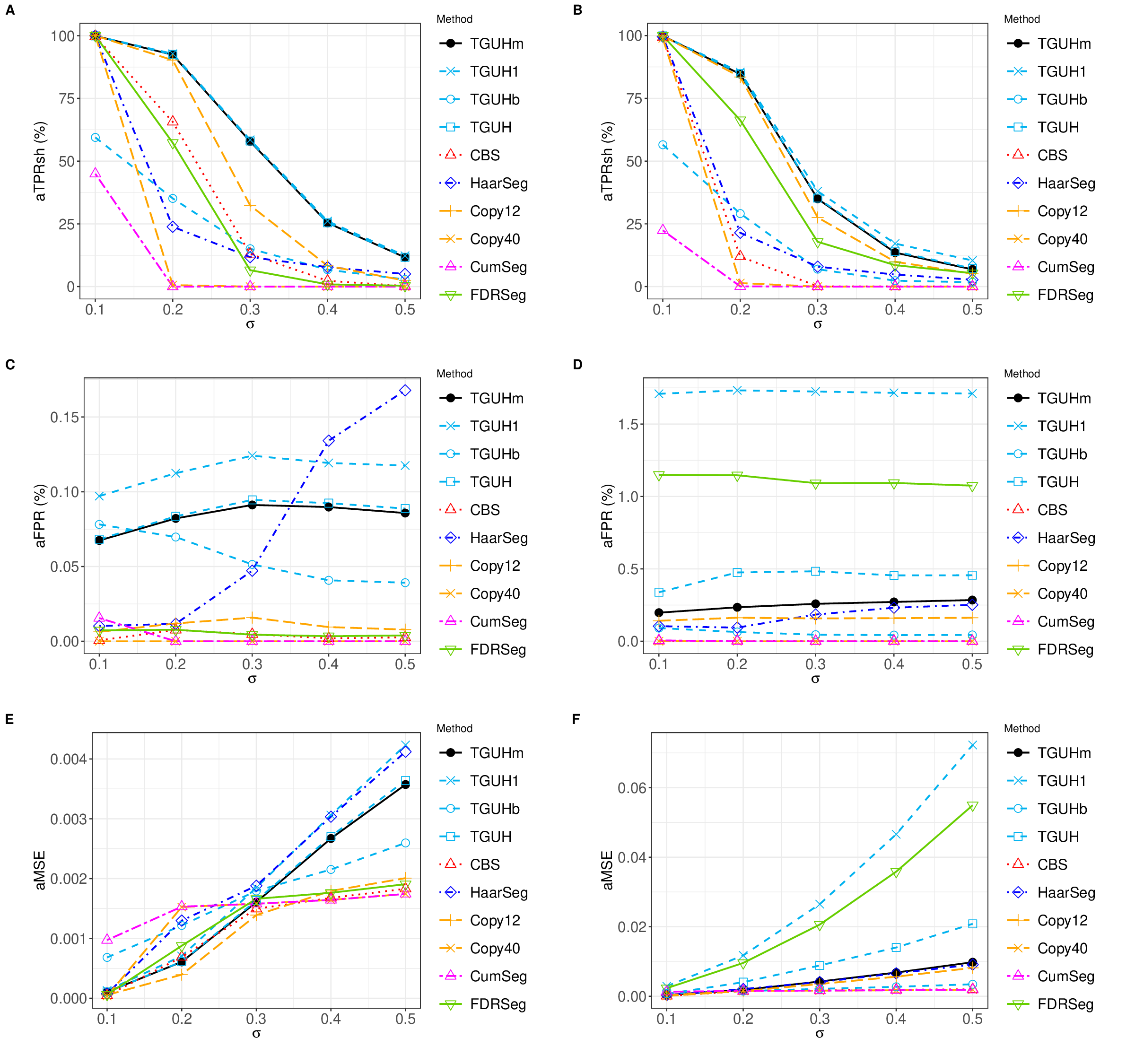}
	\end{center}
	\caption{Performance metrics for 1000 replicates of the third test function (see panel \textbf{C} of Figure~\ref{fig:test_obj}). \textbf{(A) (B)} Average of true positive rate in estimating change-points that corresponds to short segments (aTPRsh). \textbf{(C) (D)} Average of false positive rate (aFPR). \textbf{(E) (F)} Average of mean-square error (aMSE) of the estimated piecewise constant signal to the true function. The left column (panels \textbf{A},\textbf{C}, and \textbf{E}) show results for  i.i.d.\ Gaussian noise $N(0,\sigma^2)$, while the right column (panels \textbf{B},\textbf{D}, and \textbf{F}) show results for noise from a mixture of two Gaussian distributions $0.95\times N(0,\sigma^2)+0.05 \times N(0,3\sigma^2)$). The aTPR results are omitted as the simulated data only contains an isolated short segment. For a quick reminder, TGUH1 denotes both TGUH and TGUHm method with $c^*=1$ while TGUHm and TGUH denote TGUHm and TGUH method with $c^*=2$, respectively. TGUHb denotes TGUH method with a localised pruning algorithm. Copy12 and Copy40 denote CopyNumber method with $\gamma$ parameter equal to 12 and 40, respectively.
	}
	\label{fig:type_6}
\end{figure*}

Figures~\ref{fig:type_8}, \ref{fig:type_9}, and \ref{fig:type_6} show the results of the simulation study using the first, second, and third true function, respectively. The corresponding quantitative results are presented Tables~S1--S3 in the Supplementary Material. Figure~\ref{fig:type_8} indicates that for the basic Gaussian noise TGUHm, TGUH, and TGUH1 outperform the other competitors in terms of estimating both short and long segments by showing the highest aTPRsh and aTPR values for all noise levels but it comes with slightly larger aFPR and aMSE than most of the tested methods. In particular, different $c^*$ values ($c^*=1$ or $c^*=2$) do not affect the results much when the noise is standard Gaussian noise. The differences in performance due to these choices are more apparent in the case when the noise comes from the mixture of two normal distributions with different variance; see the right side plots of Figure~\ref{fig:type_8}. For aTPRsh, all of TGUHm, TGUH, and TGUH1 do not show a significant difference (still the best) but TGUH1 is marginally the worst in terms of aFPR and aMSE. On the other hand, TGUHm and TGUH are much better than TGUH1 for both aFPR and aMSE which indicate that setting $c^*$ equal to two successfully reduces spurious change-points (spikes) which are caused by the occasional extreme outliers. Moreover, compared to TGUH, TGUHm tends to return fewer false positives. This shows that adding the unconnected thresholding is preferable as it allows us to use the $c^*$ to control the minimum segment width compared to using connected thresholding alone. 

Figures~\ref{fig:type_9} and \ref{fig:type_6} show that when the simulated data only contain short segments, TGUHm still performs very well by showing excellent results in terms of aTPRsh and aMSE without excess false positives. Besides TGUHm, the TGUH1, TGUH, Copy12, and FDRSeg methods do well in terms of estimating short segments. CBS also performs well in estimating short segments for the standard Gaussian noise but it is not as good as those methods when the noise is the Gaussian mixture noise. It also shows poor performance in terms of aTPRsh when the true function only contains an isolated short segment in the middle of a very long segment as shown in Figure~\ref{fig:type_6}. The FDRSeg method, while showing good performance for short test signals, it is sensitive to occasional extremely noisy observations. This reflects in larger aFPR and aMSE for Gaussian mixture noise. 

Compare to the other methods, CumSeg and Copy40 tend to miss some change-points and fail to estimate short segments even for low level of noise contamination. This also indicates that the performance of CopyNumber method is sensitive to the selection of $\gamma$. Therefore, in practice, it may be necessary to test a number of $\gamma$ values to find the optimal one. 

\begin{figure*}[h!]
	\begin{center}
		\includegraphics[scale=0.35]{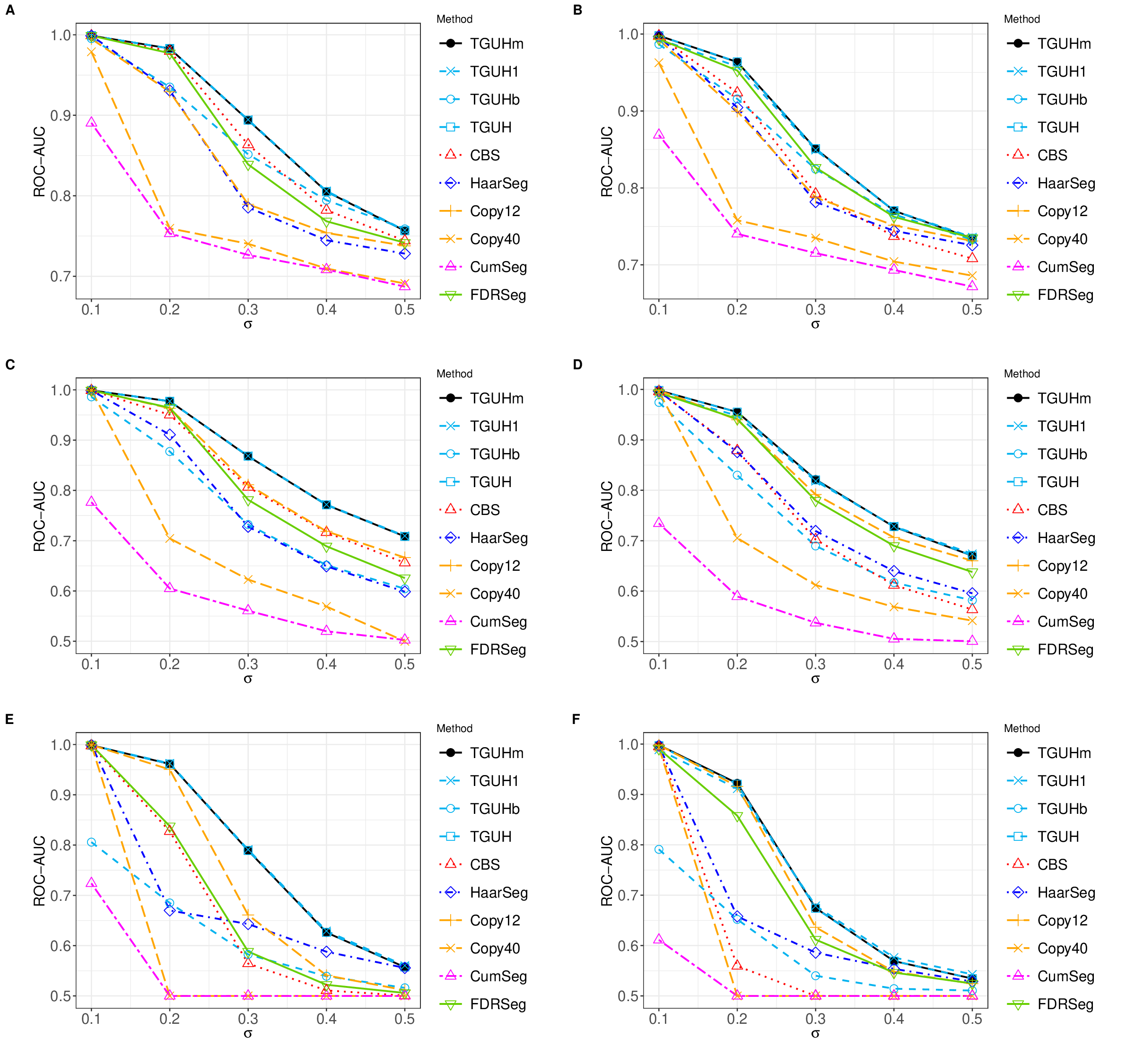}
	\end{center}
	\caption{AUC of ROC of the methods applied to the first, second, and third test functions (Figure~\ref{fig:test_obj}).  The left column (panels \textbf{A},\textbf{C}, and \textbf{E}) show results for  i.i.d.\ Gaussian noise $N(0,\sigma^2)$, while the right column (panels \textbf{B},\textbf{D}, and \textbf{F}) show results for noise from a mixture of two Gaussian distributions $0.95\times N(0,\sigma^2)+0.05 \times N(0,3\sigma^2)$). The first (panels \textbf{A} and \textbf{B}), second (panels \textbf{C} and \textbf{D}), and third (panels \textbf{E} and \textbf{F}) row correspond to the first, second, and third test function presented in Figure~\ref{fig:test_obj}, respectively. For a quick reminder, TGUH1 denotes both TGUH and TGUHm method with $c^*=1$ while TGUHm and TGUH denote TGUHm and TGUH method with $c^*=2$, respectively. TGUHb denotes TGUH method with a localised pruning algorithm. Copy12 and Copy40 denote CopyNumber method with $\gamma$ parameter equal to 12 and 40, respectively.}
	\label{fig:ROCall}
\end{figure*}

To see the trade-off between TPR and FPR, ROC curves of each of the simulations across different noise levels are considered and the corresponding area under the curve (AUC) is reported in Figure~\ref{fig:ROCall}. For both noise types used in the simulation, it is quite clear that performance in terms of AUC severely deteriorates when the noise level increases. Specifically, in a very extreme case where there is only a short altered segment in the underlying test function, the performance of CBS, Copy40, and CumSeg methods are very poor which are shown in panel E and F of Figure~\ref{fig:ROCall}. This is reflected by their AUC scores that drastically drop to $0.5$ for noise with standard deviation $\sigma$ greater than $0.1$. 

\begin{figure*}[h]
	\begin{center}
		\includegraphics[scale=0.35]{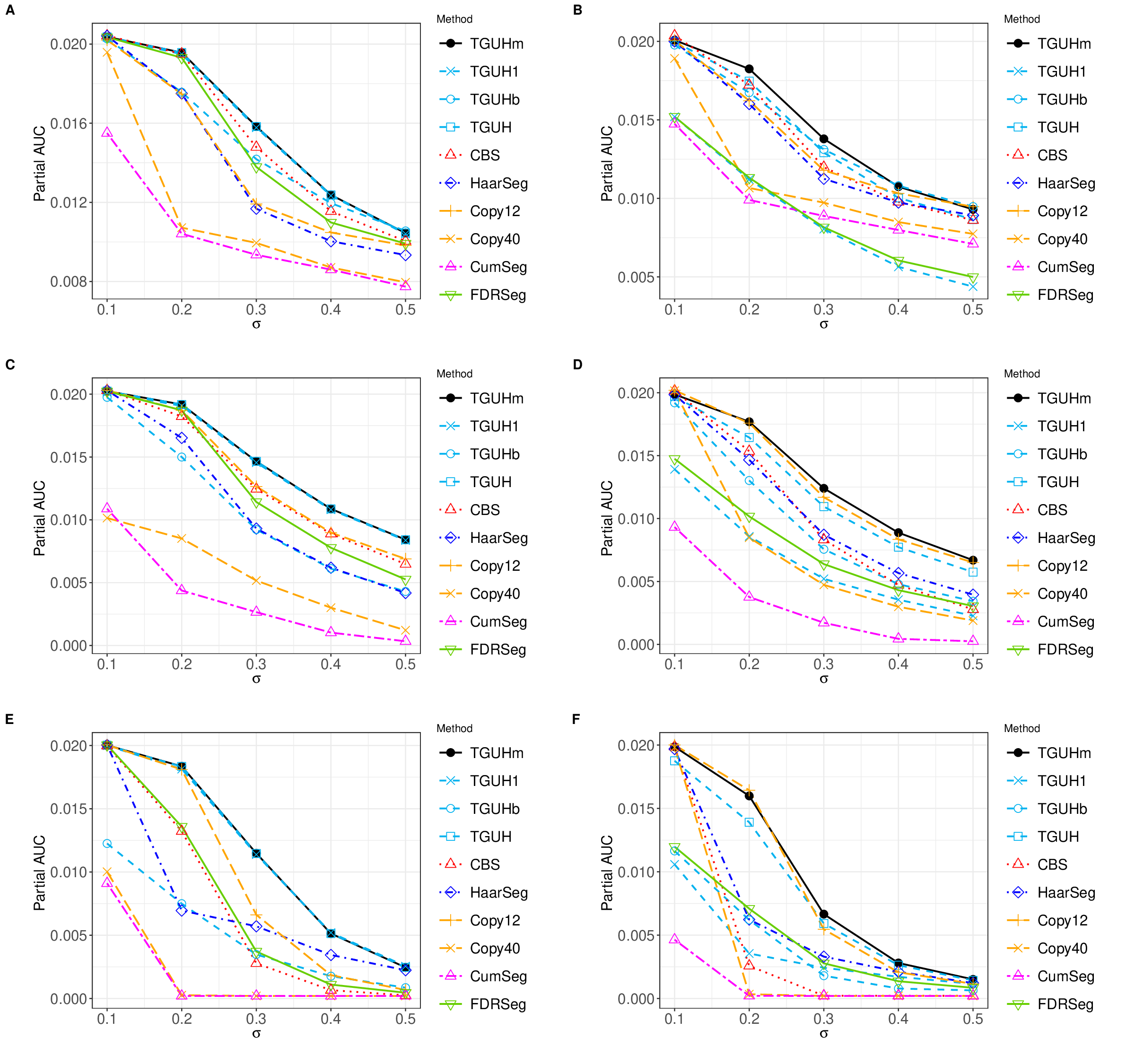}
	\end{center}
	\caption{Partial AUC $FP<20$ of ROC of the methods applied to the first, second, and third test functions (Figure~\ref{fig:test_obj}).  The left column (panels \textbf{A},\textbf{C}, and \textbf{E}) show results for  i.i.d.\ Gaussian noise $N(0,\sigma^2)$, while the right column (panels \textbf{B},\textbf{D}, and \textbf{F}) show results for noise from a mixture of two Gaussian distributions $0.95\times N(0,\sigma^2)+0.05 \times N(0,3\sigma^2)$). The first (panels \textbf{A} and \textbf{B}), second (panels \textbf{C} and \textbf{D}), and third (panels \textbf{E} and \textbf{F}) row correspond to the first, second, and third test function presented in Figure~\ref{fig:test_obj}, respectively. For a quick reminder, TGUH1 denotes both TGUH and TGUHm method with $c^*=1$ while TGUHm and TGUH denote TGUHm and TGUH method with $c^*=2$, respectively. TGUHb denotes TGUH method with a localised pruning algorithm. Copy12 and Copy40 denote CopyNumber method with $\gamma$ parameter equal to 12 and 40, respectively.
	}
	\label{fig:pROCall}
\end{figure*}

Figure~\ref{fig:ROCall} also indicates that te results of TGUHm, TGUH1, and TGUH almost overlap and has better AUC than the other methods in most of the noise levels considered. This is due to the number of false positive that corresponds to spikes being very low compared to the number of negative cases. To avoid the issues caused by this condition, Figure~\ref{fig:pROCall} shows partial AUC for early retrieval area ($FP<20$). The partial AUC results show that TGUHm method still provides excellent results over all the noise levels for both noise types considered. Moreover, TGUHm shows a significant improvement compared to the original TGUH (TGUH1, TGUH) method for the heavier-tailed noise type that caused extreme outliers in the data (See right column of Figure~\ref{fig:pROCall}). This indicates that TGUHm method is competent to reduce spurious change-points commonly found in the original TGUH method caused by extreme outliers.

\begin{figure*}[p]
	\begin{center}
		\includegraphics[scale=0.35]{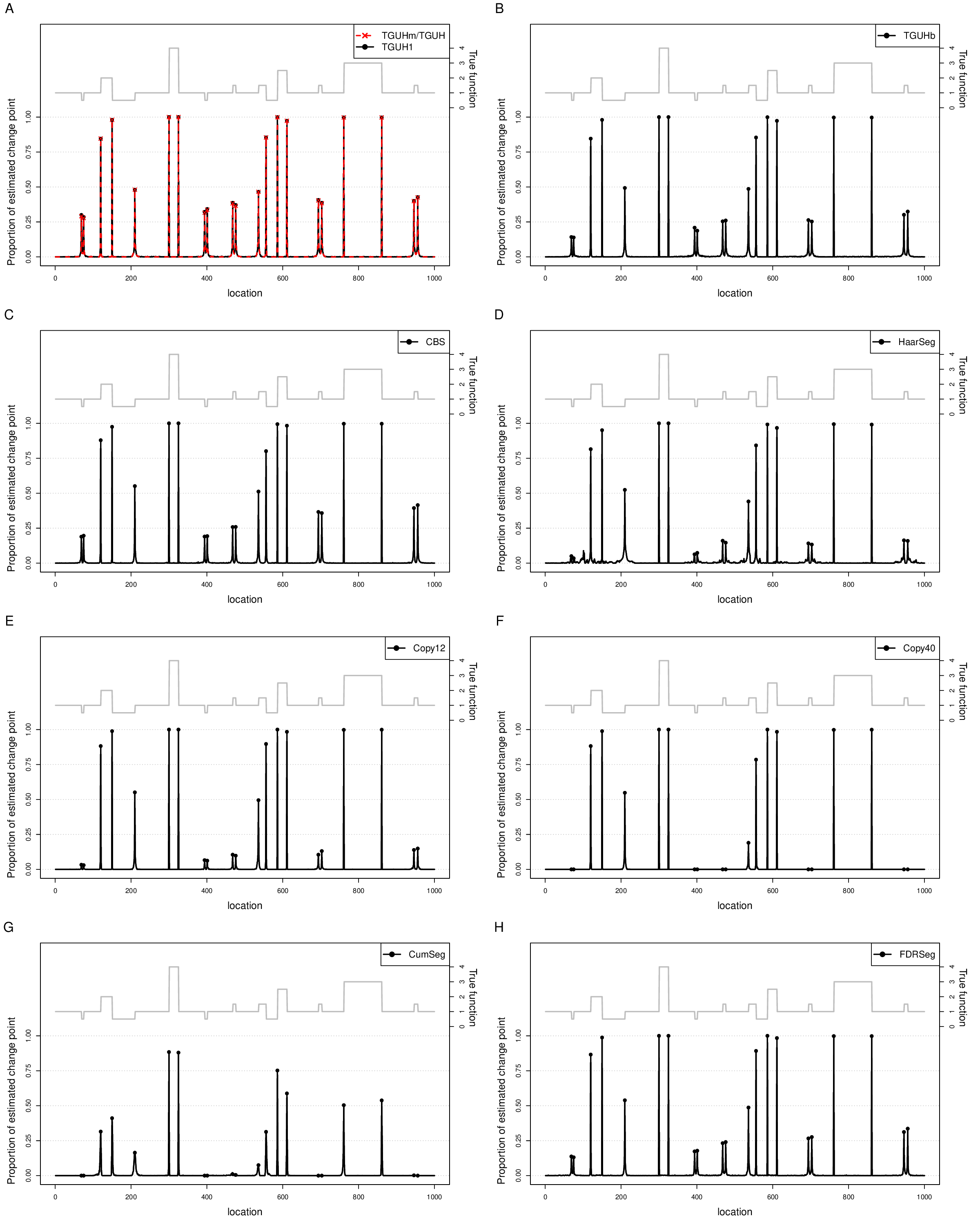}
	\end{center}
	\caption{Proportion of times a change-point is estimated against location out of 1000 simulated datasets contaminated with a mixture of two Gaussian distributions $0.95\times N(0,\sigma^2)+0.05 \times N(0,3\sigma^2)$ for $\sigma^2 = 0.3^2$. The dots denote the proportion of detection at locations where there are actual change-points. The grey solid line is the corresponding test function, repeated here from panel \textbf{A} of Figure~\ref{fig:test_obj} for ease of reference. The left and right vertical axis show the proportion of replicates where a change-point is estimated and the corresponding test function's height, respectively. For a quick reminder, TGUH1 denotes both TGUH and TGUHm method with $c^*=1$ while TGUHm and TGUH denote TGUHm and TGUH method with $c^*=2$, respectively. TGUHb denotes TGUH method with a localised pruning algorithm. Copy12 and Copy40 denote CopyNumber method with $\gamma$ parameter equal to 12 and 40, respectively.
	}
	\label{fig:frecloc8}
\end{figure*}

To investigate the performance of each method in estimating the correct location of each change-point, Figures~\ref{fig:frecloc8} show the proportion of times (from 1000 simulated datasets) that each method detects a change-point at each location along the sequence. Here, we only show the results for simulated data using the first true function which are contaminated with noise from a mixture of two normal distributions $0.95\times N(0,\sigma^2)+0.05 \times N(0,3\sigma^2)$ where $\sigma=0.3$; results for the remaining types of simulated data and basic Gaussian noise can be found in Figures~S2--S6 in the Supplementary Material. Since the results of TGUHm and TGUH almost overlap, we plotted the results of TGUHm and TGUH as one line. 

Based on Figure~\ref{fig:frecloc8}, the proposed method TGUHm has the highest sensitivity in terms of detecting short segments while still showing a relatively good performance in estimating long segments. Most of the methods have narrow `peaks' in the location of the true changes, which indicate the ability of the methods to estimate change-point exactly at the true location over 1000 iteration. But careful inspection shows that the HaarSeg method has small peaks near the true change-points. This shows the tendency of HaarSeg to produce spurious change-points near the true change-point locations. This is commonly found in Haar wavelet-based methods and is its main weakness which has successfully been overcome by our proposed method. 

%\clearpage
\subsection{Real NGS data}
\begin{figure*}[p]
	\begin{center}
		\includegraphics[scale=0.5]{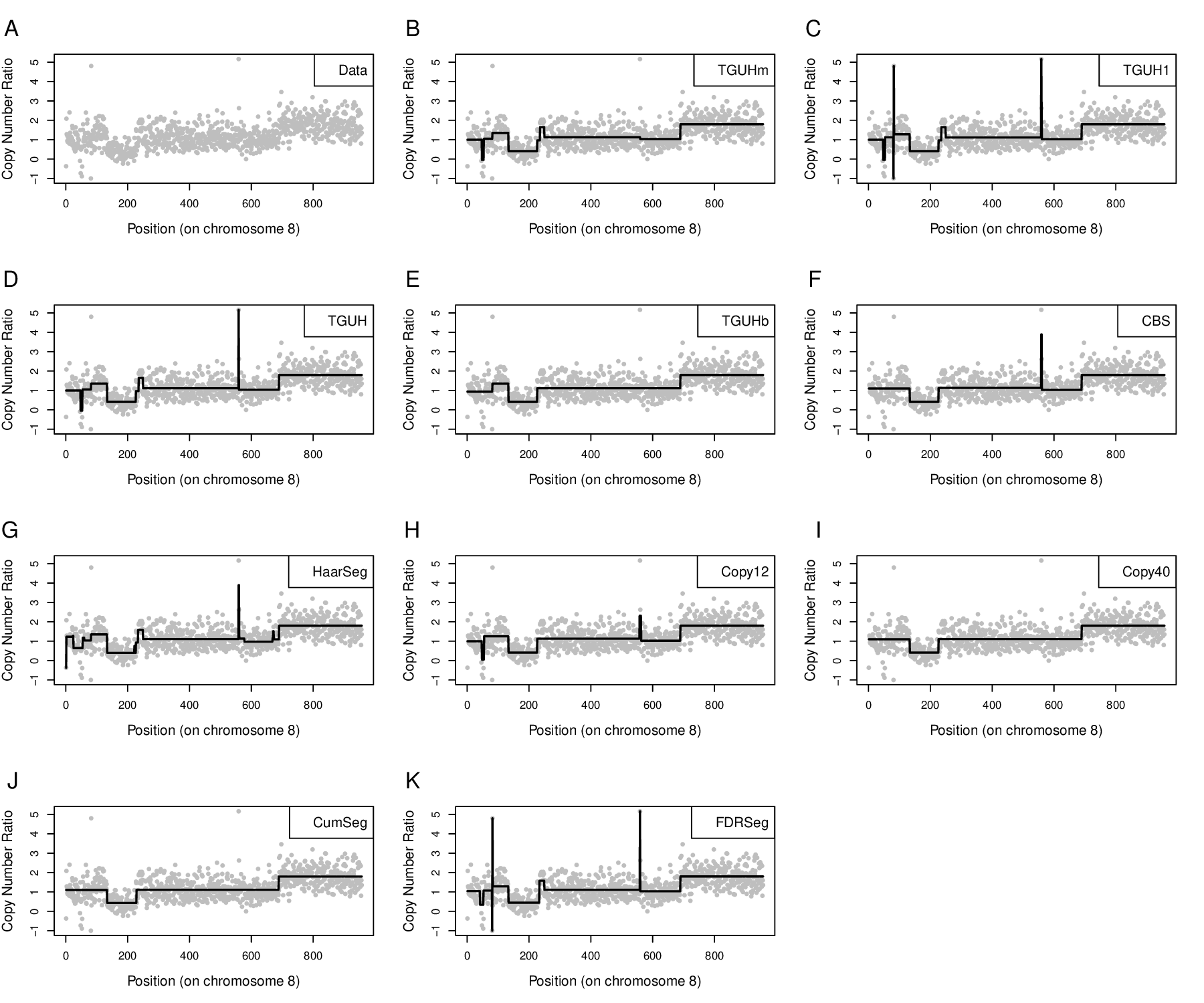}
	\end{center}
	\caption{CNA estimate as a result of segmentation of chromosome 8 in patient TMA-93. \textbf{(A)} The copy number ratio data of chromosome 8 in Patient TMA-93. \textbf{(B)} TGUHm segmentation (the detailed location of altered regions are shown on Table~\ref{tab:gene}). \textbf{(C)} TGUH1 segmentation. \textbf{(D)} TGUH segmentation. \textbf{(E)} TGUHb segmentation. \textbf{(F)} CBS segmentation, \textbf{(G)} HaarSeg segmentation, \textbf{(H)} CopyNumber segmentation with $\gamma=12$ and \textbf{(I)} $\gamma=40$, \textbf{(J)} CumSeg segmentation, and \textbf{(K)} FDRSeg segmentation. For a quick reminder, TGUH1 denotes both TGUH and TGUHm method with $c^*=1$ while TGUHm and TGUH denote TGUHm and TGUH method with $c^*=2$, respectively. TGUHb denotes TGUH method with a localised pruning algorithm. Copy12 and Copy40 denote CopyNumber method with $\gamma$ parameter equal to 12 and 40, respectively.
	}
	\label{fig:TMA-93chr8}
\end{figure*}

To illustrate the types of segments produced in more detail, Figure~\ref{fig:TMA-93chr8} presents the results of segmentation based on TGUHm, TGUH, TGUH1, TGUHb, CBS, HaarSeg, CumSeg, and FDRSeg method in chromosome~8 of patient~TMA-~93 \cite{Belvedere2012} (see Figure~S8 of the Supplementary Material for segmentation of the whole genome and Figure~S7 and S9 for segmentation of other patients). Figure~\ref{fig:TMA-93chr8} shows that most of the methods except CBS, Copy40, TGUHb and CumSeg estimate short segments loss at the position around 50. We know from the simulation study that CBS, CumSeg, TGUHb And Copy40 are not sensitive to short segments compared to the remaining methods. This indicates that there may be short altered segments at that region with high probability. TGUHm estimates more short segments than Copy12 but less than HaarSeg and FDRSeg which corresponds to results from the simulation, where we have seen that Copy12 is less sensitive to short segments while both HaarSeg and FDRSeg tend to form more false positives than TGUHm. This feature is found not only in NGS data but also in aCGH data as presented in Figure~S10--S12 of the Supplementary Material.

In more detail, Table~\ref{tab:gene} shows the list of altered regions on TGUHm CNA estimate of chromosome 8 patient~TMA-~93. TGUHm method indicates loss at position $48$--$52$ covers around 1Mb from 7,05Mb to 7,8Mb of patient~TMA-~93 chromosome 8. It has been known that some of DEFA and DEFB genes are maps within those locations \citep{ensembl2022}. The DEFA and DEFB or $\alpha$ and $\beta$ defensins are small cationic antimicrobial peptides which contribute to kill bacteria, fungi, and enveloped viruses by disruption of the microbial membrane in phagocytes, the skin, and the mucosa (including that of the lung) \citep{HIEMSTRA20067}. It means that the estimated short altered segment indicates deletion of DEFA and DEFB in TMA-~93 patient. There is also MYC gene that is maps in altered region between 106,2Mb -- 146,25Mb. Previous study from \cite{baykara2015amplification} has shown that the amplification MYC gene highly correlated to lung cancer.

\begin{table}
	\centering
	\caption{List of altered regions on TGUHm CNA estimate of chromosome 8 Patient~TMA-~93 (which corresponds to panel B of Figure~\ref{fig:TMA-93chr8}). }
	{\begin{tabular}{lc} \toprule \\ 
			Aberration & Location on chromosome 8   \\ \midrule
			Deletion & 7,05Mb -- 7,8Mb  \\
			Amplification & 12,15Mb -- 19,8Mb\\
			Deletion & 19,95Mb -- 35,0Mb \\
			Amplification & 35,25Mb -- 37,2Mb \\
			Amplification & 106,2Mb -- 146,25Mb\\
			\bottomrule
	\end{tabular}}
	\label{tab:gene}
\end{table} 

In the context of squamous cell carcinoma lung cancer, several subsequent studies including analysis across samples are needed to investigate the association between lung cancer and the indicated altered segments. Even though our method only can be used for analyze CNA per sample, its segmentation result still holds a crucial role to provide possible genes candidate to be analysed further.

Moreover, in this example, the differences between TGUH and TGUHm are clearly seen. We can identify spikes (due to extreme single points) produced by TGUH1 ($c^*=1$) which are completely removed by TGUHm ($c^*=2$). While the standard TGUH without the unconnected thresholding could not remove all those spikes, even when we set $c^*=2$. Since we do not know the truth in real data, it is difficult to confirm whether the spikes are real changes or not although, given that they are single points, we expect a priori that they are not. However, these results indicate that one should consider TGUHm when it is appropriate to assume a value for the minimum segment length $c^*$. 

\section{Discussion and Concluding Remarks}

In this article, we have adapted the TGUH method for use with copy number data by modifying its thresholding technique so that it is no longer constrained to the  `unary-binary tree' structure; we call this adaptation the TGUHm method. This modification aims to address the tendency of the standard TGUH method to overestimate CNA `spikes' when there are isolated points with extremely high CNA ratios as frequently seen in real data. By  modifying the thresholding procedure, the TGUHm method is shown to be successful in reducing those spikes. Our experience suggests that setting $c^*$ to two gives the most benefit in terms of reducing the occurrences of single-point spikes in the segmented CNA. When $c^*$ is increased further to three, four, or even five, say, the results are very similar to $c^*=2$ as shown in Figure~S13--15 of the Supplementary Material. This indicates that, for reasonably low values of $c^* \geq 2$, the conclusion is not sensitive to the choice of $c^*$. Therefore, if users do not generally know the minimum length of expected altered segments, we suggest $c^*=2$ as it provides a significant improvement to the performance of the method. We speculate that the need for a high $c^*$ value may arise in a case with correlated noise.

Our results also suggest that the method has good operating characteristics to detect segments of different sizes. Some methods may have a tendency to identify more short segments or long segments. The proposed methods demonstrably work well for both short and long segments. This result becomes increasingly crucial in the case of low-coverage NGS data such as is the case in this study. This is because, for example, a 1 Mb segment is represented by only 5-7 windows or data points \citep{Gusnanto2014}. In this case, segmentation methods are tested to the limit of detection, and the choice of the method becomes crucial. In the context of high-coverage NGS data, then the same 1 Mb segment can be represented by hundreds of points. In such cases, we expect that most of the segmentation methods will perform well with very little difference between their results. 

It is interesting to note that, in general, the main advantage of the wavelet approach is its ability to decompose data and represent them as a series of coefficients at several scales. This enables us to scrutinize the data variability at different scales and identify the wavelet coefficients that belong to the noise terms. Moreover, the proposed method is able to adjust the location of the jumps in the unbalanced Haar wavelets to match the likely structure of the signal at hand which result in more precise change-point location estimates than alternatives based on traditional balanced Haar wavelets. This advantage has made the proposed method viable for CNA segmentation.

Even though it has been shown that the proposed methods perform well in identifying change-points where the data contain  Gaussian noise with constant variance, analysing data with more complex noise structures is still challenging. This is a subject for future research and more work will be needed to further improve the TGUHm method in this setting. Lastly, as shown in the Supplementary Material, it is worth noting that the proposed TGUHm method is of general use in terms of CNA data produced by different technologies such as aCGH arrays.

\section*{Acknowledgments}
The first author (MAU) wishes to acknowledge the financial support of a School of Mathematics PhD Scholarship at the University of Leeds.

\bibliographystyle{unsrtnat}
\bibliography{references}  %%% Uncomment this line and comment out the ``thebibliography'' section below to use the external .bib file (using bibtex) .

%%% Uncomment this section and comment out the \bibliography{references} line above to use inline references.
% \begin{thebibliography}{1}

% 	\bibitem{kour2014real}
% 	George Kour and Raid Saabne.
% 	\newblock Real-time segmentation of on-line handwritten arabic script.
% 	\newblock In {\em Frontiers in Handwriting Recognition (ICFHR), 2014 14th
% 			International Conference on}, pages 417--422. IEEE, 2014.

% 	\bibitem{kour2014fast}
% 	George Kour and Raid Saabne.
% 	\newblock Fast classification of handwritten on-line arabic characters.
% 	\newblock In {\em Soft Computing and Pattern Recognition (SoCPaR), 2014 6th
% 			International Conference of}, pages 312--318. IEEE, 2014.

% 	\bibitem{hadash2018estimate}
% 	Guy Hadash, Einat Kermany, Boaz Carmeli, Ofer Lavi, George Kour, and Alon
% 	Jacovi.
% 	\newblock Estimate and replace: A novel approach to integrating deep neural
% 	networks with existing applications.
% 	\newblock {\em arXiv preprint arXiv:1804.09028}, 2018.

% \end{thebibliography}

\end{document}